\documentclass[runningheads]{llncs}

\usepackage[utf8]{inputenc}
\usepackage{multirow}

{}

\newcommand{\tr}[3]{\tau_{#1,q_{#2} \rightarrow q_{#3}}}{}
\newcommand{\trs}[3]{T_{#1,q_{#2} \rightarrow q_{#3}}}{}
\newcommand{\enstrs}[3]{{\mathcal T}_{#1,q_{#2} \rightarrow q_{#3}}}{}

\newcommand{\delt}[3]{\delta_{#1,q_{#2} \rightarrow q_{#3}}}{}
\newcommand{\Sp}{S^{+}}{}
\newcommand{\Sm}{S^{-}}{}
\newcommand{\ders}[3]{D_{{#1},{q_{#2}},q_{#3}}}{}
\newcommand{\dere}[3]{D^{*}_{{#1},{q_{#2}},q_{#3}}}{}

{}

\usepackage{xcolor}
\usepackage{colortbl}

\usepackage{graphicx}
\begin{document}
\title{Improved SAT models for NFA learning}
\author{Frédéric Lardeux\orcidID{0001-8636-3870} 
\and
Eric Monfroy\orcidID{0001-7970-1368} }
\authorrunning{F. Lardeux et al.}
\institute{LERIA, University of Angers, France
\email{firstname.lastname@univ-angers.fr}}
\maketitle              % typeset the header of the contribution
\begin{abstract}
Grammatical inference is concerned with the study of algorithms for learning automata and grammars from words. We focus on learning Nondeterministic Finite Automaton of size $k$ from samples of words. To this end, we formulate the problem as a SAT model. The generated SAT instances being enormous, we propose some model improvements, both in terms of the number of variables, the number of clauses, and clauses size. These improvements significantly reduce the instances, but at the cost of longer generation time. We thus try to balance instance size vs. generation and solving time. We also achieved some experimental comparisons and we analyzed our various model improvements.

\end{abstract}

%%%%%%%%%%%%%%%%%%%%%%%%%%%%%%%%%%%%%%%%%%%%%%%%%%%%%%%%%%%%%%%%%%%%%%
\section{Introduction}

Grammatical inference~\cite{ColinBook} is concerned with the study of algorithms for learning automata and grammars from words.  It plays a significant role in numerous applications, such as compiler design, bioinformatics,  speech recognition, 
%syntactic and structural 
pattern recognition, machine learning, and others. The problem we address in this paper is learning a finite automaton from samples of words $S=\Sp \cup \Sm$, which consist of positive words ($\Sp$) that are in the language and must be accepted by the automaton, and negative words ($\Sm$) that must be rejected by the automaton. A non deterministic automaton (NFA) being generally a smaller description for a language than an equivalent deterministic automaton (DFA),  we focus here on NFA inference. An NFA is represented by a 5-tuple $(Q , \Sigma , \Delta, q_1, F )$ where $Q$ is a finite set of states, the vocabulary $\Sigma$ is a finite set of symbols, the transition function $\Delta  : Q \times \Sigma \rightarrow {\mathcal P}(Q)$ associates a set of states to a given state and a given symbol, $q_1 \in Q$ is the initial state, and $F \subseteq Q$ is the set of final states.

Not to mention DFA (e.g.,~\cite{HeuleMarijn2013Smsu}), the problem for NFA has been tackled from a variety of angles. In~\cite{WieczorekBook} a wide panel of techniques for NFA inference is given. 
Some works focus on the design of ad-hoc algorithms, such as \textit{DeLeTe2}~\cite{delete2} that is based on state merging methods. More recently, a new family of algorithms for regular languages inference was given in~\cite{DBLP:conf/wia/PargaGR06}. 
%Some work consists in solving the problem with some metaheuristic-based solvers. For example, in~\cite{tomita82}, the author applies hill-climbing to modify finite automata to accept a regular language. 
Some approaches are based on metaheuristic, such as in~\cite{tomita82} where hill-climbing is applied in the context of regular language, or~\cite{DBLP:conf/icgi/Dupont94} which is based on genetic algorithm.
%for grammatical inference.
In contrast to metaheuristics, complete solvers are always able to find a solution if there exists one, to prove the unsatisfiablility of the problem, and to find the optimal solution in case of optimization problems. In this case, generally, the problem is modeled as a Constraint Satisfaction Problem (CSP~\cite{Rossi2006}). For example, in~\cite{WieczorekBook}, an Integer Non-Linear Programming (INLP) formulation of the problem is given. Parallel solvers for minimizing the inferred NFA size  are presented in~\cite{jastrzab2016,jastrzab2017}. The author of~\cite{jastrzab2019} proposes two strategies, based on variable ordering, for solving the CSP formulation of the problem.

In this paper, we are not interesting in designing or improving a solver, but we focus in improving models of the problem in order to obtain faster solving times using a standard SAT solver. Modeling is the process of translating a problem into a CSP consisting in decision variables and constraints linking these variables. 
The INLP model for NFA inference of~\cite{WieczorekBook} cannot be easily modified to reduce the instances: to our knowledge, 
%among all the improvements we propose in this paper, 
only Property~\ref{proplambda} of our paper could be useful for the INLP model, and we do not see any other possible improvement. 
We thus start with a rather straightforward conversion of the INLP model %of~\cite{WieczorekBook} 
into the propositional satisfiablity problem (SAT~\cite{Garey1979}). This is our base SAT model to evaluate our improvements.
%
%We thus start from a conversion into the propositional satisfiablity problem (SAT~\cite{Garey1979}) of the INLP model of~\cite{WieczorekBook}: this will be our base model to evaluate our improvements. This conversion is rather straightforward, converting INLP variables into Boolean variables, and INLP constraints into conjunctions and disjunctions of Boolean variables. 
%Each state is associated to a variable which is true if the state is final. The existence of a transition between two states reading a symbol is represented by a Boolean variable. A sequence of transitions in the NFA is represented by a \textit{c\_transition}, i.e., a conjunction of Boolean variables. Sequences of transitions induce disjunctions that must be transformed into a Conjunctive Normal Form (CNF) in order to obtain a formula suitable for the majority of SAT solvers.
%
The model, together with a training sample,
%of positive and negative words, 
lead to a SAT instance that we solve with a standard SAT solver. The generated SAT instances are very huge: the order of magnitude is $|S|.(|\omega|+1).k^{|\omega|}$ clauses, where $k$ is the number of states of the NFA, $\omega$ is the longest word of $S$,
% the training sample, 
and $|S|$ is the number of words of the training sample. 
We propose three main improvements to reduce the generated SAT instances. The first one prevents generating 
subsumed constraints. Based on a multiset representation of words, the second one avoid generating some useless constraints. The last one is a weaker version of the first one, based on prefixes of words. 
The first improvement returns smaller instances than the second one, which in turn returns smaller instances than the third one. However, the first improvement is very long and costly, whereas the third one is rather fast. We are thus interested in balancing generation and solving times against instance sizes.
%(which may also create some memory problems).
%
%When possible, we evaluate the size of the SAT instances (numbers of variables and clauses, length of clauses), and the gain we have with the model improvements. 
%
We achieved some experiments with the Glucose solver~\cite{glucose} to compare the generated SAT instances. 
The results show that our improvements are worth: larger instances could be solved, and faster. Generating the smallest instances can be too costly, and the best results are obtained with a good balance between instance sizes and generation/solving time.

%\smallskip

This paper is organized as follows. In Section~\ref{sec:modeling}, we describe the problem and we give the basic SAT model. We also evaluate the size of the generated instances.
Section~\ref{sec:improvement} presents 3 model improvements, together with sketches of algorithms to generate them. Section~\ref{sec:expe} exposes our experimental results and some analysis. We finally conclude in Section~\ref{sec:conclusion}.

%autres approches: asp en 2020, 

%%%%%%%%%%%%%%%%%%%%%%%%%%%%%%%%%%%%%%%%%%%%%%%%%%%%%%%%%%%%%%%%%%%%%%
\section{Modeling the problem in SAT} \label{sec:modeling}

The non-linear integer programming (INLP) model of~\cite{WieczorekBook,jastrzab2017}
%In~\cite{WieczorekBook,jastrzab2017}, a non-linear integer programming model is given. This INLP model 
cannot be easily improved or simplified. Indeed, the only improvement proposed in~\cite{WieczorekBook} and~\cite{jastrzab2017} corresponds to Property~\ref{proplambda} (given in the next section). In this section, we thus present a SAT formulation of the NFA inference problem. This SAT model permits many improvements to reduce the size of the generated SAT instances.

%%%%%%%%%%%%%%%%%%%%%%%%%%%%%%%%%%%%%%%%%%%%%%%%%%
%\subsection{The problem: NFA inference}
\paragraph{\textbf{The NFA inference problem}}
Consider
an alphabet $\Sigma=\{s_1,\ldots,s_n\}$ of $n$ symbols;
    a training sample $S=\Sp \cup \Sm$, where $\Sp$ (respectively $\Sm$) is a set of \textit{positive words} (respectively \textit{negative words}) from $\Sigma^{*}$;
    and an integer $k$.   
The problem consists in building a NFA of size $k$ which validates words of $\Sp$, and rejects words of $\Sm$. 
%We say that $\Sp$ is a set of "positive" words, and $\Sm$ is a set of "negative" words.
The problem can be extended to an optimization problem: it consists in inferring a minimal NFA for $S$, i.e., an NFA minimizing $k$. However, we do not consider optimization in this paper.

%%%%%%%%%%%%%%%%%%%%%%%%%%%%%%%%%%%%%%%%%%%%%%%%%%
\paragraph{\textbf{Notations}}
Let $A=(Q,\Sigma, q, F)$ be a NFA with:
 $Q=\{q_1,\ldots,q_k\}$ a set of states,
    $\Sigma$ a finite alphabet (a set of symbols),
    $q$ the initial state,
    and $F$ the set of final states.
The symbol 
$\lambda$ represents the empty word. We denote by $K$ the set $\{1, \ldots,k\}$.
A transition from $q_j$ to $q_k$ with the symbol $s_i$ is denoted by $\tr{s_i}{j}{k}$. Consider the word  $w=w_1 \ldots  w_n$ with $w_1, \ldots, w_n$ in $\Sigma$. Then, the notion of transition is extended to $w$ by $\trs{w}{i_1}{i_{n+1}}$ which is a sequence of transitions $\tr{w_1}{i_1}{i_2}$, \ldots, $\tr{w_n}{i_{n}}{i_{n+1}}$. The set of candidate transitions for $w$ between the states $q_{i_1}$ and $q_{i_l}$ in a NFA of size $k$ is $\enstrs{w}{i_1}{i_l}=\{\trs{w}{i_1}{{i_l}} ~|~ \exists i_2, \ldots i_{i_l-1}\in K, ~ \trs{w}{i_1}{{i_l}} = \tr{w_1}{{i_1}}{{i_2}},  \ldots, \tr{w_l}{{i_{l}-1}}{{i_l}} 
    \}$.

%%%%%%%%%%%%%%%%%%%%%%%%%%%%%%%%%%%%%%%%%%%%%%%%%%
\paragraph{\textbf{A SAT model}}
Our base model is a conversion into SAT of the nonlinear integer programming problem given in~\cite{WieczorekBook} or~\cite{jastrzab2017}.
Consider the following variables:
\begin{itemize}
    \item $k$ the size of the NFA we want to build,
    \item $F=\{f_1, \ldots, f_k\}$ a set of $k$ Boolean variables determining whether states $q_1$ to $q_k$ are final or not,
    \item and $\Delta=\{\delt{s}{i}{j}| s \in \Sigma \textrm{~and~} i,j \in K\}$ a set of $n.k^2$ variables determining whether there is or not a transition $\delt{s}{i}{j}$, i.e., a transition from state $q_i$ to state $q_j$ with the symbol $s$, for each $q_i$, $q_j$, and $s$.
\end{itemize}    
A transition $\trs{w_1 \ldots  w_n}{i_1}{i_{n+1}}=\tr{w_1}{i_1}{i_2}$,  \ldots $\tr{w_n}{i_{n}}{i_{n+1}}$ exists if and only if the conjunction 
$d=\delt{w_1}{i_1}{i_2} \wedge  \ldots \wedge \delt{w_n}{i_{n}}{i_{n+1}}$ is true. 
We call $d$ a c\_transition, and we say that $d$ models $\trs{w_1 \ldots  w_n}{i_1}{i_{n+1}}$. 
We denote by $\ders{w}{i}{j}$ the set of all c\_transitions for the word $w$ between states $q_i$ and $q_j$.
\smallskip

The problem can be modeled with 3 sets of equations:
\begin{enumerate}
% lambda
    \item If the empty word $\lambda$ is in $\Sp$ or in $\Sm$, we can determine whether the first state is final or not:
    \begin{eqnarray}
    \textrm{if } \lambda \in \Sp, ~~~~~~f_1 \label{lambda1} \\
    \textrm{if } \lambda \in \Sm, ~~~~    \neg f_1 \label{lambda2}
    \end{eqnarray}    
    
%positive word
    \item For each word $w \in \Sp$, there is at least a transition starting in $q_1$ and ending in a final state $q_j$:
    \begin{eqnarray}
    %\bigwedge_{w \in \Sp} \left[
     \bigvee_{j \in K} \bigvee_{~d \in \ders{w}{1}{j}} \big( d \wedge f_j \big) \label{m1}
    %\right]
    \end{eqnarray}
    
    With the Tseitin transformations~\cite{Tseitin1983}, we create one auxiliary variable for each combination of a word $w$, a state $j \in K$, and a transition $d \in \ders{w}{1}{j}$: 
    \begin{eqnarray}
    aux_{w,j,d} \leftrightarrow d \wedge f_j \nonumber
    \end{eqnarray}
    For each $w$, we obtain a formula in CNF:
    \begin{eqnarray}
    %\bigwedge_{w \in \Sp} 
    \bigwedge_{j \in K} 
    \bigwedge_{~d \in \ders{w}{1}{j}}
    \left[
    (\neg aux_{w,j,d} \vee (d \wedge f_j)) %\wedge  (\neg aux_{w,j,d} \vee f_j) 
    \right]  \label{aux1Mk} \\
    %
    %\bigwedge_{w \in \Sp} 
    \bigwedge_{j \in K} 
    \bigwedge_{~d \in \ders{w}{1}{j}}
    (aux_{w,j,d} \vee \neg d \vee \neg f_j)   \label{aux2Mk} \\
    %
     %\bigwedge_{w \in \Sp} 
    \bigvee_{j \in K} 
    \bigvee_{~d \in \ders{w}{1}{j}}
        aux_{w,j,d} \label{aux3Mk}
    \end{eqnarray}
    $d$ is a conjunction, and thus $\neg aux_{w,j,d} \vee d$ is a conjunction of $|w|$ binary clauses: $(\neg aux_{w,j,d} \vee \delt{w_1}{1}{{i_2}}) \wedge  \ldots \wedge (\neg aux_{w,j,d} \vee \delt{w_{|w|}}{i_{|w|}}{{i_{|w|+1}}})$. 
    
    $|\ders{w}{1}{j}|=k^{|w|-1}$ since for each symbol of $w$ there is $k$ possible moves in the NFA, except for the last symbol which leads to $q_j$. 
    Thus, we have  $(|w|+1).k^{|w|}$ binary clauses for Constraints~(\ref{aux1Mk}), $k^{|w|}$ $(|w|+2)$-ary clauses for Constraints~(\ref{aux2Mk}), and one $k^{|w|}$-ary clause for Constraints~(\ref{aux3Mk}). We have added $k^{|w|}$ auxiliary variables.

    % negative sample
    \item For each $w \in \Sm$ and each state $q_j$, either there is no complete transition from state $q_1$ to $q_j$, or $q_j$ is not final:
    \begin{eqnarray}
    \neg \left[
     \bigvee_{j \in K} \bigvee_{~d \in \ders{w}{1}{j}} \big(d  \wedge f_j \big) 
    \right] 
    \label{negM}
    \end{eqnarray}
    Constraints (\ref{negM}) are already in CNF, and we have $k^{|w|}$ $(|w+1|)$-ary clauses.
    \end{enumerate}

    Thus, the constraint model $M_k$ for building a NFA of size $k$ is: 
    \[M_k=   \bigwedge_{w \in \Sp} 
    \Big((\ref{aux1Mk}) \wedge 
    (\ref{aux2Mk}) \wedge 
    (\ref{aux3Mk}) \Big) 
    \wedge 
    \bigwedge_{w \in \Sm} (\ref{negM})\]
    and is possibly completed by $(\ref{lambda1})$ or $(\ref{lambda2})$ if $\lambda \in \Sp$ or $\lambda \in \Sm$.

%%%%%%%%%%%%%%%%%%%%%%%%%%%%%%%%%%%%%%%%%%%%%%%%%%
\paragraph{\textbf{Size of the models}}

Considering $\omega_{+}$, the longest word of $\Sp$, and $\omega_{-}$, the longest word of $\Sm$, the number of constraints in model $M_k$ is bounded by:
    \begin{itemize}
        \item $|\Sp|.(|\omega_{+}|+1).k^{|\omega_{+}|}$ binary clauses;
        \item $|\Sp|.k^{|\omega_{+}|}$ $(|\omega_{+}|+2)$-ary clauses;
        \item $|\Sp|$ $k^{|\omega_{+}|}$-ary clauses;
        \item $|\Sm|.k^{|\omega_{-}|}$ $(|\omega_{-}|+1)$-ary clauses.
    \end{itemize}
    
    The number of Boolean variables is bounded by:
    \begin{itemize}
        \item $k$ variables in $F$ determining final states;
        \item $n.k^2$ variables determining existence of transitions;
        \item $|\Sp|.k.^{|\omega_{+}|}$ auxiliary variables $aux_{w,j,d}$.
    \end{itemize}

It is thus obvious that it is important to improve the model $M_k$.

%%%%%%%%%%%%%%%%%%%%%%%%%%%%%%%%%%%%%%%%%%%%%%%%%%%%%%%%%%%%%%%%%%%%%%
\section{Improving the SAT model} \label{sec:improvement}

We now give some properties that can be used for improving the SAT model. By abuse of language, we will say that a model $M_1$ is smaller than a model $M_2$ whereas we should say that the SAT instance generated with $M_1$ and data $D$ is smaller than the instance generated with $M_2$ and $D$.
A first and simple improvement is based on the following property.

\begin{property}[Empty word $\lambda$] \label{proplambda}
If $\lambda \in \Sm$, then each c\_transition ending in $q_1$ does not have to be considered when generating the constraints related to  the word $w \in S$.
\end{property}
Indeed, if $w$ is positive, it cannot be accepted by a transition ending in $q_1$; similarly, if $w$ is negative, $\neg d \vee \neg  f_1$ is always true.
When $\lambda \in \Sp$, the gain is not very interesting: $f_1$ can be omitted in Constraints~(\ref{negM}), (\ref{aux1Mk}), and (\ref{aux2Mk}). This does not really reduce the instance, and a standard solver would simplify it immediately. 

%\medskip

Whereas a transition is an ordered sequence, the order of conjuncts in a c\_transition is not relevant, and equal conjuncts can be deleted. Thus, a c\_transi\-tion may model several transitions, and may correspond to several words. By abuse of language, we say that a c\_transition  ends in a state $q_j$ if it corresponds to at least a transition ending in $q_j$. Thus, a c\_transition may end in several states. 
We consider an order on c\_transitions. Let $d$ and $d''$ be two c\_transitions. Then, $d \preceq d''$ if and only if there exists a c\_transition $d'$ such that $d \wedge d'=d''$. In other words, each transition variable $\delt{s}{i}{j}$ appearing in $d$ also appears in $d''$.
This order is used in the two first model improvements which are based on c\_transitions. The third model improvement is based on transitions.
We now consider some redundant constraints.
\begin{property}[Redundant constraints] \label{propredundant}
When a state $q_i$ cannot be reached, each outgoing transition becomes free (it can be assigned true or false), and $q_i$ can be final or not. 
In order to help the solver, all the corresponding variables can be assigned an arbitrary value.
For each state $q_j$, $j \not = 1$:
\[
\big( \bigwedge_{i \in K, i \not = j} \bigwedge_{s \in \Sigma} \neg \delt{s}{i}{j} \big)
\rightarrow
 \neg f_j \wedge \big( \bigwedge_{i \in K} \bigwedge_{s \in \Sigma} \neg \delt{s}{j}{i} \big)
\]
In CNF, these constraints generate (for all $q_j$), $(k-1).(k.n+1)$ redundant clauses of size $n.(k-1)+1$.
\end{property}
These constraints are useful when looking for a NFA of size $k$ when $k$ is not the minimal size of the NFA. Compared to SAT instance size, these redundant constraints can be very helpful without being too heavy.

%\medskip

Note that in our implementation, for all the models, we always simplify instances using Property~\ref{proplambda} and removing duplicate transition variables in c\_transitions (i.e., $\delt{s}{i}{j} \wedge \ldots \wedge \delt{s}{i}{j}$ is simplified into $\delt{s}{i}{j} \wedge \ldots$). Moreover, we also generate the redundant constraints as defined in Property~\ref{propredundant}.

%%%%%%%%%%%%%%%%%%%%%%%%%%%%%%%%%%%%%%%%%%%%%%%%%%
\paragraph{\textbf{Improvement based on c\_transitions subsumption.}}

This first improvement consists in removing tautologies for negative words, and some constraints and unsatisfiable disjuncts for positive words.

\begin{property}[c\_transition subsumption] \label{propssmot}
Let $v$ be a negative word from $\Sm$, and $\neg d_v \vee \neg q_j$ be a Constraint~(\ref{negM}) generated for the c\_transition $d_v$ for $v$ ending in state $q_j$. We denote this constraint $c_{v,d_v,q_j}$.
Consider a positive word $w$ from $\Sp$, and $d_w$ a c\_transition for $w$ ending in $q_j$ such that $d_v \preceq d_w$. Then, each $d_w \wedge f_j$ will be false due to $c_{v,d_v,q_j}$. Thus, Constraints~(\ref{aux1Mk}) and (\ref{aux2Mk}) corresponding to $w$, $d_w$, and $q_j$ will force to satisfy $\neg aux_{w,j,d_w}$; hence, they can be omitted and $aux_{w,j,d_w}$ can be removed from Constraints~(\ref{negM}).
Similarly, consider $\omega$ from $\Sm$, and $d_{\omega}$ a c\_transition for $w$ ending in $q_j$ such that $d_v \preceq d_w$. Then, Constraint~(\ref{negM}), $\neg d_v \vee \neg q_j$, will always be true (due to the constraint $c_{v,d_v,q_j}$), and can be omitted.
\end{property}

%We cannot compute the size of the reduced SAT instance in the general case. However, w
We can compute the size of the reduced SAT instance when the smaller word is a prefix. 
Let $v \in \Sm$ and $w \in S$ be words such that $w=v.v'$, i.e., $v \subseteq w$ and $v$ is a prefix of $w$. Then, using Property~\ref{propssmot}: if $w \in \Sm$, the number of clauses generated for $w$ is reduced to $(k-1).k^{|w|-1}$ clauses of size $|w+1|$; if $w \in \Sp$, the number of clauses generated for $w$ is reduced to
$(|w|+1).(k-1).k^{|w|-1}$ binary clauses for Constraints~(\ref{aux1Mk}), 
$(k-1).k^{|w|-1}$ $(|w|+2)$-ary clauses for Constraints~(\ref{aux2Mk}), 
and one clause of size $(k-1).k^{|w|-1}$ for Constraint~(\ref{aux3Mk}). The number of auxiliary variables is reduced to  $(k-1).k^{|w|-1}$.

%\medskip
Operationally, we have a two step mechanism. First, for each negative word, each c\_transition together with its ending state is generated and stored in a database of couples (c\_transition, ending state) that we call c\_couple. Then, for generating constraints for a word $w$, each of its c\_couple is compared to the database. If a c\_transition for $w$ ending in $q_j$ is smaller than a c\_transition from the database also ending in $q_j$, then the corresponding constraints are not generated, as shown above.
We call $M_{k,all}$ this reduced model.

%%%%%%%%%%%%%%%%%%%%%%%%%%%%%%%%%%%%%%%%%%%%%%%%%%
\paragraph{\textbf{Improvement based on Multisets.}}

Although efficient in terms of generated instance sizes, the previous improvement is very costly in memory and time. It becomes rapidly intractable.
This second improvement also uses Property~\ref{propssmot}. It is a weakening of the above operational mechanism that does not omit every subsumed c\_transition. This mechanism is less costly. Hence, generated instances will be a bit larger, but the balance generation time against instance size is very good. 
The idea is to order words in order to search in a very smaller database of c\_couples (c\_transition, ending state) when generating constraints for a word $w$. Moreover, this order will also imply the order for generating constraints.

We associate each word to a multiset which support is the vocabulary $\Sigma$. The word $w$, is thus associated with the multiset $ms(w)=\{s_1^{|w|_{s_1}}, \ldots, s_n^{|w|_{s_n}}\}$ where $|w|_{s_i}$ is the number of occurrences of the symbol $s_i$ in $w$. Note that several words can have the same multiset representation.
Based on multiset inclusion ($\{s_1^{{a'_1}}, \ldots, s_n^{{a'_n}}\} \subseteq_{{\mathcal M}} \{s_1^{{a_1}}, \ldots, s_n^{{a_n}}\} \Leftrightarrow \forall i, a'_i \leq a_i$), we can now define the notion of word inclusion, noted $\subseteq_{\omega}$. Consider $w$ and $w'$, two words of $\Sigma^{*}$, then:
\[
w' \subseteq_{\omega} w \Leftrightarrow ms(w') \subseteq_{{\mathcal M}} ms(w)
\]
    
Consider a sample $S=\Sp \cup \Sm$. Let $\top(S)$ be the multiset defined as 
\[
\top(S)=\{s_1^{{1+max_{w \in (S)}\{|w|_{s_1}\}}}, \ldots, s_n^{{1+max_{w \in (S)}\{|w|_{s_n}\}}}\}
\]
and $\bot=\{s_1^{{0}}, \ldots, s_n^{{0}}\}$. Then, $\top(S)$ represents words which are not in the sample $S$, and $\bot$ represents the empty word $\lambda$ which may be in $S$. %In the following, we will omit the sample name, and write $\top$ instead of $\top(S)$.

Consider the sample $S=\Sp \cup \Sm$. Let $MS(S)=\{ms(w) | w \in \Sp \cup \Sm\}$ be the set of the representations of words of $S$. Then, $(MS(S) \cup \{\bot,\top(S)\}, \subseteq_{\mathcal M})$ is a lattice.
Let $m$ be a multiset of $MS(S)$. Then, $inf(m)$ is the set of multisets $\{m' \in MS(S) ~|~ m' \subseteq_{{\mathcal M}} m \}$.
This lattice of multisets defines the data structure used for constraint generation. For generating constraint of a word $w$ of a multiset $m$, we now only compare its c\_couples with the database of c\_couples of words $w' \in \Sm$ with $w' \subseteq_{\omega} w$, i.e., words represented by multisets smaller than $m$.

\smallskip

The negative words that allow to reduce the most, are the ones represented by the smallest multiset. We thus also propose a mechanism to reduce the database (c\_transition, ending state) with the most useful c\_couples, i.e., the ones from smallest words. Let $level(m)$ be the "level" of the multiset defined by: $level(m)=0$ if $m=\bot$, $1+max_{m' \in inf(m)}(level(m'))$ otherwise.
Given a multiset $m$, and a threshold $l$, the $base$ function returns all the multisets $m'$ of level smaller than $l$, and such that $m' \subseteq_{\mathcal M} m$:
$base(p,l)= 
 \{n \in inf(p)  ~|~ level(n)\leq l \} \bigcup \big(\bigcup_{p' \in inf(p)} base(p',l)\big)$ if $p \not = \bot$,
 $\emptyset$ otherwise.

Based on Property~\ref{propssmot}, c\_couples of the negative words of these multisets will be used to reduce constraint generation of the words of $m$.
We call this model $M_{k,mset,l}$, with $l$ a given threshold.
If $base$ is called with the threshold 0, the database will be empty and the complete instance will be generated: $M_{k,mset,0}=M_k$. If $base$ is called with the maximum level of the lattice, then, the database will be the largest one built with all the smaller words, and we will thus obtain the smallest instances with this notion of lattice. However, the larger the threshold, the longer the generation time, and the smaller the SAT instance. 
With the maximal threshold, the generated instances will be a bit larger than with the previous improvement ($M_{k,all} \subseteq M_{k,mset,max}$), but the generation is significantly faster.
%
%\medskip
For lack of space, we cannot give here the complete algorithms for generating this improved model.

%The complete algorithms are given in Annex.

%%%%%%%%%%%%%%%%%%%%%%%%%%%%%%%%%%%%%%%%%%%%%%%%%%
\paragraph{\textbf{Improvements based on Prefixes.}}

Although faster to generate, the second model is still costly. We now propose a kind of weakening of Property~\ref{propssmot}, restricting its use to prefix.

\begin{property}[Prefix] \label{propprefix}
Let $w \in S$ be a word from the sample. Consider $\dere{w}{i}{j}$ the set of c\_transitions defined by:
\[
\dere{w}{i}{j}=
%\bigvee_{l \in K \atop l \not = j} 
\bigvee_{l \in K, l \not = j} 
    \Bigg( 
        \bigg( 
            \bigvee_{d_u \in \dere{u}{i}{l}} d_u   
        \wedge 
        \Big( 
            \bigvee _{d_v \in \dere{v}{l}{i}} d_v 
        \Big)
        \bigg)
    \Bigg)    
\]
if  $w=u.v$, and $u \in \Sm$; otherwise, $\dere{w}{i}{j}=\ders{w}{i}{j}$.
Then, 
\[
    \forall d \in \ders{w}{i}{j} \setminus \dere{w}{i}{j}, \neg d \vee \neg f_j
\]
\end{property}    
Hence, this property allows us to directly generate the reduced constraints, for negative or positive words, without comparing c\_couples  with a database.

Let $w=u_1 \ldots u_n$ be a  word from $S$ such that $u_1 \in \Sm$, $u_1.u_2 \in \Sm$, and $u_1 \ldots u_{n-1} \in \Sm$ and for each $i < n$, there does not exist a decomposition $u_i=u'_i.u''_i$ such that $u_1 \ldots u_{i-1}.u'_i \in \Sm$. 
Then, if $w \in \Sp$, using several times Property~\ref{propprefix}, Constraints~(\ref{aux1Mk}), (\ref{aux2Mk}), and (\ref{aux3Mk}) can be replaced by Constraints~(\ref{aux1Mk2}), (\ref{aux2Mk2}), and (\ref{aux3Mk2}) where $l_0=q_1$ and $N=[1, \ldots, n]$: 
%in Figure~\ref{cpref}.
\begin{eqnarray}
    \bigwedge_{
                            i \in N, 
                            l_i \in K \setminus \{l_j| 1\leq j < i\}~
              }
    \bigwedge_{
                    ~i \in N, 
                    d_i \in \ders{u_i}{l_i-1}{l}
              }  
    \left[
    (\neg aux_{w,l_1,\ldots,l_n} \vee (d_1 \wedge \ldots \wedge d_n \wedge f_j))
    %\wedge  (\neg aux_{w,l_1,\ldots,l_n} \vee f_j) 
    \right]  \label{aux1Mk2}   ~~\\
    \bigwedge_{
                    i \in N, 
                    l_i \in K \setminus \{l_j| 1\leq j < i\}~
              }
    \bigwedge_{
                    ~i \in N, 
                    d_i \in \ders{u_i}{l_i-1}{l}
              }  
    (aux_{w,l_1,\ldots,l_n}  \vee \neg d_1 \vee \ldots \vee \neg d_n \vee \neg f_j)   \label{aux2Mk2}   ~~\\
    \bigvee_{
                    i \in N, 
                    l_i \in K \setminus \{l_j| 1\leq j < i\}~
              }
    \bigvee_{
                    ~i \in N, 
                    d_i \in \ders{u_i}{l_i-1}{l}
              }  
    aux_{w,l_1,\ldots,l_n} \label{aux3Mk2}  ~~  ~~
\end{eqnarray}

Similarly, if $w \in \Sm$, using several times Property~\ref{propprefix}, Constraints~(\ref{negM}) can be replaced by Constraints~(\ref{negM2}):
%as shown in Figure~\ref{cpref}.
\begin{eqnarray}
    \bigwedge_{
                    i \in N, 
                    l_i \in K \setminus \{l_j| 1\leq j < i\}~
              }
    \bigwedge_{
                    ~i \in N, 
                    d_i \in \ders{u_i}{l_i-1}{l}
              }  
   (\neg d_1 \vee \ldots \vee \neg d_n \vee \neg f_j)   \label{negM2} 
    \end{eqnarray}
The number of clauses and variables generated for $w \in \Sp$ is reduced to:
\begin{itemize}
    \item $(|w|+1).\big( \prod_{i=1}^{n}(k-i+1) \big). k^{|w|-n}$ binary clauses for Constraints~(\ref{aux1Mk2}), 
    \item $\big( \prod_{i=1}^{n}(k-i+1) \big). k^{|w|-n}$ $(|w|+2)$-ary clauses for Constraints~(\ref{aux2Mk2}), 
    \item one clause of size $\big( \prod_{i=1}^{n}(k-i+1) \big)$ for Constraint~(\ref{aux3Mk2}),
    \item and the number of auxiliary variables is reduced to  $\big( \prod_{i=1}^{n}(k-i+1) \big)$.
\end{itemize}
For $w \in \Sm$, Constraints (\ref{negM2}) are already in CNF and they correspond    
    to $\big( \prod_{i=1}^{n}(k-i+1) \big). k^{|w|-n}$ $(|w+1|)$-ary clauses.    
Interestingly, these new counts of clauses (and more especially the factor $k-i+1$ with $i=n$) also give us a lower bound for $k$: $k$ must be greater than or equal to $n$, the number of nested prefixes in a word. 
This new improved model, that we call $M_{k,pref}$, is not much larger than $M_{k,mset}$, but it is significantly faster to generate.

%%%%%%%%%%%%%%%%%%%%%%%%%%%%%%%%%%%%%%%%%%%%%%%%%%
\paragraph{\textbf{Improvement order.}} \label{model_order}

We have defined various models for inference of NFA of size $k$ that can be ordered by their sizes:
$
M_{k,all} \subseteq M_{k,mset,l\_max} \subseteq m_{k,pref} \subseteq M_k
$.
Note that $M_{k,mset,l}$ with $l \not = l\_max$, and $M_{k,pref}$ cannot be compared in the general case; their sizes depend on the instance, the number and size of prefixes, and on the given level $l$. 
In the next section, we compare these models not only in terms of instance size, but also in terms of generation and resolution time.

%%%%%%%%%%%%%%%%%%%%%%%%%%%%%%%%%%%%%%%%%%%%%%%%%%%%%%%%%%%%%%%%%%%%%%
\section{Experimental results} \label{sec:expe}
We suspect that, with respect to their generation time, the models are in reverse order of the order given above. Thus, we are interested in findng  the best balance between three parameters: model size v.s. generation time + SAT solving time.

%\paragraph{Experimental context}
The experiments were carried out on a computing cluster with Intel-E5-2695 CPUs and 128 GB of memory. Running times were limited to 2 hours for the generation of SAT instances, and 3 hours to solve them. We used the Glucose~\cite{glucose} SAT solver with the default options.%(based on the well-known Minisat~\cite{minisat} solver) was used with the default options.
% 
%\paragraph{Benchmarks based on the StaMinA competition}
The benchmarks are based on the training set of the StaMinA Competition (http://stamina.chefbe.net).  We selected 12 instances\footnote{We conserved the "official" name used during the Stamina Competition.} with a sparsity $s \in \{12.5\%, 25\%, 50\%, 100\%\}$ and an alphabet size $|\Sigma| \in \{2, 5, 10\}$. For each of them, we limited the number of words to $|\Sp|=|\Sm|=10$ and 20 for a maximal size of words equal to 7 and to $|\Sp|=|\Sm|=20$  for a maximal size of words equal to 10. We generate CNF instances for different NFA sizes ($k \in\{3,4,5\}$). Consequently, we obtained 96 instances.

%\paragraph{Analysis}
Table \ref{tab-results} presents a synthetic view of our experiments. The 4 first columns detail the instances: size of the NFA ($k$), size of the longest word ($|\omega|$), number of positive (and negative) words ($|\Sp|$), and the model. The next columns provide average values over the 12 instances for the modeling time ($T_{Model}$), the number of variables ($\#Var$), the number of clauses ($\#Cl$), the solving time ($T_{solve}$), and the total modeling+solving time  ($T_{total}$). We do not indicate the standard deviations but they are very close to zero. "-" indicates that no result was obtained before the time-out.
\renewcommand{\arraystretch}{1}
\setlength{\tabcolsep}{5pt}
\begin{table}[!ht]
\centering 
\caption{Comparison on 96 generated instances between the models $m_{k,all}$, $m_{k,mset,l_{max}}$, $m_{k,mset,1}$, $m_{k,mset,3}$, and $m_{k,pref}$. Instances are grouped by size of the NFA ($k$), size of the longest word ($|\omega|$), and number of positive (and negative) words ($|\Sp|$). For each line, obtained values are average on 12 instances.}
\label{tab-results}
{\scriptsize
\begin{tabular}{|c|c|c|c|r r r r r|}
\hline
k&$|\omega|$&$|S^+|$&Model&$T_{model}$&\#Var.&\#Cl.&$T_{solve}$&$T_{total}$\\
\hline
\multirow{18}{*}{3}   	&	  \multirow{12}{*}{7} 	&	  \multirow{6}{*}{10}    	&	 $m_k$ 	&	 0.19 	&	6742	&	61366	&	0.22	&	0.41	\\
 	&	  	&	    	&	 $m_{k,all}$ 	&	 0.68 	&	4310	&	37789	&	0.14	&	0.82	\\
 	&	  	&	    	&	 $m_{k,mset,l_{max}}$ 	&	 0.17 	&	4742	&	42020	&	0.14	&	0.31	\\
 	&	  	&	    	&	 $m_{k,mset,1}$ 	&	 0.18 	&	5517	&	49484	&	0.16	&	0.34	\\
 	&	  	&	    	&	 $m_{k,mset,3}$ 	&	 0.17 	&	4822	&	42850	&	0.14	&	0.31	\\
 	&	  	&	    	&	 $m_{k,pref}$ 	&	 0.18 	&	6466	&	58645	&	0.2	&	0.38	\\
\cline{3-9}																	
	&		&	  \multirow{6}{*}{20}   	&	 $m_k$ 	&	 0.48 	&	14830	&	134302	&	1.58	&	2.06	\\
 	&	  	&	    	&	 $m_{k,all}$ 	&	 2.62 	&	8274	&	72569	&	1.64	&	4.26	\\
 	&	  	&	    	&	 $m_{k,mset,l_{max}}$ 	&	 0.42 	&	8929	&	79030	&	1.22	&	1.64	\\
 	&	  	&	    	&	 $m_{k,mset,1}$ 	&	 0.45 	&	11179	&	99811	&	1.39	&	1.84	\\
 	&	  	&	    	&	 $m_{k,mset,3}$ 	&	 0.46 	&	9148	&	81188	&	1.27	&	1.73	\\
 	&	  	&	    	&	 $m_{k,pref}$ 	&	 0.43 	&	13689	&	123390	&	1.71	&	2.14	\\
\cline{2-9}																	
 	&	   \multirow{6}{*}{10}  	&	 \multirow{6}{*}{20}   	&	 $m_k$ 	&	11	&	303519	&	3276974	&	397.68	&	408.68	\\
 	&	  	&	    	&	 $m_{k,all}$ 	&	 746.08 	&	108417	&	1172093	&	79.98	&	826.06	\\
 	&	  	&	    	&	 $m_{k,mset,l_{max}}$ 	&	 9.87 	&	122423	&	1313463	&	143.32	&	153.19	\\
 	&	  	&	    	&	 $m_{k,mset,1}$ 	&	 9.04 	&	208610	&	2255307	&	233.97	&	243.01	\\
 	&	  	&	    	&	 $m_{k,mset,3}$ 	&	 9.06 	&	134720	&	1443357	&	156.24	&	165.3	\\
 	&	  	&	    	&	 $m_{k,pref}$ 	&	 8.88 	&	281408	&	3040802	&	270.04	&	278.92	\\
\hline																	
\multirow{18}{*}{4}     	&	 \multirow{12}{*}{7}  	&	 \multirow{6}{*}{10} 	&	    $m_k$ 	&	 1.46 	&	45014	&	428775	&	10.3	&	11.76	\\
 	&	  	&	    	&	 $m_{k,all}$ 	&	 19.42 	&	32956	&	302835	&	5.59	&	25.01	\\
 	&	  	&	    	&	 $m_{k,mset,l_{max}}$ 	&	 1.64 	&	35362	&	328938	&	5.58	&	7.22	\\
 	&	  	&	    	&	 $m_{k,mset,1}$ 	&	 1.42 	&	39242	&	369600	&	7.12	&	8.54	\\
 	&	  	&	    	&	 $m_{k,mset,3}$ 	&	 1.56 	&	36048	&	336637	&	5.58	&	7.14	\\
 	&	  	&	    	&	 $m_{k,pref}$ 	&	 1.3 	&	43655	&	414141	&	10.69	&	11.99	\\
\cline{3-9}																	
 	&	   	&	\multirow{6}{*}{20}    	&	 $m_k$ 	&	 3.93 	&	100984	&	950473	&	83.55	&	87.48	\\
 	&	  	&	    	&	 $m_{k,all}$ 	&	 93.48 	&	64428	&	588293	&	74.55	&	168.03	\\
 	&	  	&	    	&	 $m_{k,mset,l_{max}}$ 	&	 4.33 	&	68041	&	628400	&	43.08	&	47.41	\\
 	&	  	&	    	&	 $m_{k,mset,1}$ 	&	 3.65 	&	83463	&	777005	&	32.32	&	35.97	\\
 	&	  	&	    	&	 $m_{k,mset,3}$ 	&	 4.27 	&	70720	&	653396	&	41.36	&	45.63	\\
 	&	  	&	    	&	 $m_{k,pref}$ 	&	 3.37 	&	94829	&	887943	&	55.88	&	59.25	\\
\cline{2-9}																	
 	&	  \multirow{6}{*}{10} 	&	 \multirow{6}{*}{20}    	&	 $m_k$ 	&	 187.59 	&	4670833	&	53350566	&	2084.78	&	2272.37	\\
 	&	  	&	    	&	 $m_{k,all}$ 	&	 - 	&	-	&	-	&	-	&	-	\\
 	&	  	&	    	&	 $m_{k,mset,l_{max}}$ 	&	 919.56 	&	2304788	&	26010946	&	651	&	1570.56	\\
 	&	  	&	    	&	 $m_{k,mset,1}$ 	&	 173.82 	&	3336332	&	38121787	&	658.7	&	832.52	\\
 	&	  	&	    	&	 $m_{k,mset,3}$ 	&	 375.34 	&	2345238	&	26693196	&	107.13	&	482.47	\\
 	&	  	&	    	&	 $m_{k,pref}$ 	&	 162.45 	&	4405201	&	50260648	&	1331.92	&	1494.37	\\
\hline																	
\multirow{12}{*}{5}     	&	 \multirow{12}{*}{7}  	&	 \multirow{6}{*}{10} 	&	    $m_k$ 	&	 6.61 	&	201651	&	1962754	&	215.06	&	221.67	\\
 	&	  	&	    	&	 $m_{k,all}$ 	&	 232.47 	&	161828	&	1526044	&	51.82	&	284.29	\\
 	&	  	&	    	&	 $m_{k,mset,l_{max}}$ 	&	 14.38 	&	169816	&	1619550	&	171.92	&	186.3	\\
 	&	  	&	    	&	 $m_{k,mset,1}$ 	&	 7.24 	&	182445	&	1759734	&	180.98	&	188.22	\\
 	&	  	&	    	&	 $m_{k,mset,3}$ 	&	 10.76 	&	172660	&	1653301	&	210.1	&	220.86	\\
 	&	  	&	    	&	 $m_{k,pref}$ 	&	 6.26 	&	196894	&	1908623	&	176.12	&	182.38	\\
\cline{3-9}																	
 	&	   	&	 \multirow{6}{*}{20}    	&	 $m_k$ 	&	 19.37 	&	456976	&	4382919	&	1268.18	&	1287.55	\\
 	&	  	&	    	&	 $m_{k,all}$ 	&	 1158.5 	&	320689	&	2995308	&	631.14	&	1789.64	\\
 	&	  	&	    	&	 $m_{k,mset,l_{max}}$ 	&	 44.01 	&	333799	&	3148787	&	1115.9	&	1159.91	\\
 	&	  	&	    	&	 $m_{k,mset,1}$ 	&	 20.24 	&	398074	&	3784691	&	1192.49	&	1212.73	\\
 	&	  	&	    	&	 $m_{k,mset,3}$ 	&	 32.82 	&	348339	&	3288509	&	1309.17	&	1341.99	\\
 	&	  	&	    	&	 $m_{k,pref}$ 	&	 16.54 	&	434008	&	4141453	&	1203.36	&	1219.9	\\
\hline
\end{tabular}
\vspace{-6mm}
}
\end{table}
From Table~\ref{tab-results}, we can draw some general conclusions about model improvements. 
As expected, $M_{k,all}$ always returns the smallest instances, and also the instances that Glucose solve the fastest. However, the generation time of these instances is very long. Thus, the total CPU time, i.e., generation + solving, is not the best. We can also see that when we increase the maximum length of words, this model does not permit to generate the instances in less than 2 hours (e.g., Table~\ref{tab-results}, for $k=4$, $\omega=10$, and $|\Sp|=20$). This model is thus tractable, but only for small instances, with short words and small samples.

$M_{k,mset,l_{max}}$ generates instances a bit larger than $M_{k,all}$. Consider the negative word $v=aaab$, and the positive word $w=ba$. $M_{k,all}$ uses some c\_transitions of $v$ to ignore some clauses of $w$ that $M_{k,mset,l_{max}}$ will not detect. For example, a loop on $aaa$ from $v$ with the same transition in $v$ is used in $M_{k,all}$ but not in $M_{k,mset,l_{max}}$. However, with the multiset  data structure, we obtain a much faster generation of instances. The total time is thus more interesting with $M_{k,mset,l_{max}}$ than with $M_{k,all}$. The generation time of $M_{k,mset,l_{max}}$ is still very high, and its interest is not always significant. For large instances, not presented in the table, $M_{k,mset,l_{max}}$ could not be generated in less than 2 hours.

For $M_{k,pref}$, we can see that the generation time becomes reasonable, and much smaller than with the two previous improvements. Although smaller than with $M_k$, the instances are larger than with $M_{k,mset,l_{max}}$. In various experiments, this improvement was the best for the total time. Note also that our training samples are not so big, and that the number of prefixes is not so important. With larger $|\Sp|$, for a fixed $k$, we should obtain better performances of $M_{k,pref}$.

We also tried two more improvements of $M_{k,mset,l}$ with $l \in \{1,3\}$. The generation time of these models is logically faster than the ones of $M_{k,mset,l_{max}}$; as planned, the SAT instances are also larger. However, we were pleasantly surprised by the total time which is much better than for $M_{k,mset,l_{max}}$. The three models $M_{k,pref}$, $M_{k,mset,1}$, and $M_{k,mset,3}$ are very difficult to compare. Depending on the instance, on the number and size of prefixes, on multiset inclusion, one can be better than the other. But for all the instances we tried, one of this 3 models was always the best of the 6 models, and they were better than $M_k$. 
\begin{table}[t]
\centering 
\caption{Focus on 2 specific instances.}
\label{tab-spec}
{\scriptsize
\begin{tabular}{|c|c|c|c|rrrrr|}
\hline
k&$|\omega|$&$|S^+|$&Model&$T_{model}$&\#Var.&\#Cl.&$T_{solve}$&$T_{total}$\\
\hline
\hline
\multicolumn{9}{|c|}{25\_training}\\
\hline
\multirow{6}{*}{5}   	&	  \multirow{6}{*}{7} 	&	  \multirow{6}{*}{20}    	&	 $m_k$ 	&	 16.72 	&	378030	&	3748314	&934.92	&	951.64	\\ 

 	&	  	&	    	&	 $m_{k,all}$ 	&	 854.47 	&	271338	&	2626880	&	841.22	&	1695.69	\\   

 	&	  	&	    	&	 $m_{k,mset,l_{max}}$ 	&	 48.71 	&	275331	&	2678349	&	1538.06	&1586.77	\\  

 	&	  	&	    	&	 $m_{k,mset,1}$ 	&	 14.25 	&	280899	&	2733709	&	895.92	&	910.17	\\ 

 	&	  	&	    	&	 $m_{k,mset,3}$ 	&	 23.67 	&	277359	&	2696089	&	1147.41	&	1171.08	\\ 

 	&	  	&	    	&	 $m_{k,pref}$ 	&	 11.76	&	338880	&	3377124	&	687.79	&	699.55	\\

\hline
\hline
\multicolumn{9}{|c|}{35\_training}\\
\hline
\multirow{6}{*}{4}   	&	  \multirow{6}{*}{10} 	&	  \multirow{6}{*}{20}    	&	 $m_k$ 	&	 163.10 	&	5253332	&	59504339	&	-	&	-	\\  
 	&	  	&	    	&	 $m_{k,all}$ 	&	 - 	&	-	&	-	&	-	&	-	\\
 	&	  	&	    	&	 $m_{k,mset,l_{max}}$ 	&	 676.22 	&	4234500	&	47661301	&	2322.42	&	2998.64	\\
 	&	  	&	    	&	 $m_{k,mset,1}$ 	&	 209.86 	&	4969772	&56092438		&	-	&	-	\\
% ATTENTION. PAS LOGIQUE	&	  	&	    	&	 $m_{k,mset,3}$ 	&	 720.46 	&	4235847	&	47676793	&	-	&	-	\\
 	&	  	&	    	&	 $m_{k,pref}$ 	&	 184.56 	&	5253332	&	59504339	&	7145.62	&	7330.18	\\
\hline
\end{tabular}
}
\vspace{-5mm}
\end{table}
%
%
%\medskip
%
%\medskip
%
%\medskip
Table \ref{tab-spec} presents a focus on 2 specific instances (25\_training and 35\_training, both with $|\Sigma|=5$) with a fixed value for k, $|\omega|$, and $|S^+|$. The columns correspond exactly to those of Table \ref{tab-results}.
%
%We now detail two instances with $|\Sigma|=5$ (Table~\ref{tab-spec}). 
For the first instance, we  clearly see the order presented in Section~\ref{model_order} for instance sizes of improved models. We can also see the reverse order in terms of generation time. When $|\Sigma|$ is small, the probability of having prefixes is higher than with larger vocabularies, and for this instance, $M_{k,pref}$ returns the best instance in terms of generation+solving time.
For the second instance, $M_{k,all}$ could not be generated in less than 2 hours. $M_k$ and $M_{k,mset,3}$ could be generated rather quickly, but could not be solved. $M_{k,pref}$ was even faster for generating the SAT instance. However, we see that there was not prefix in the training set (the size of instances of $M_k$ and $M_{k,pref}$ are the same). The overhead for taking prefixes into account is rather insignificant (12\% of generation time). Since the solving time was close to the timeout, the $M_k$ instance did not succeed to be solved while the $M_{k,pref}$ instance succeeded (the small difference of 55 s., i.e., less than 0,8 \%, is certainly due to clause order in the SAT instance). This instance shows that $M_{k,mset,l_{max}}$ can be the best model in terms of total time. This is due to the fact that there is no negative word being prefix of another word from $S$, and that the lattice is rather "wide", with a long branch. Hence, $M_{k,mset,l}$ is interesting when $l$ is large for this training sample.

%%%%%%%%%%%%%%%%%%%%%%%%%%%%%%%%%%%%%%%%%%%%%%%%%%%%%%%%%%%%%%%%%%%%%%
\section{Conclusion} \label{sec:conclusion}
In the context of grammatical inference, we proposeed various model improvements for learning Nondeterministic Finite Automaton of size $k$ from samples of words. Our base model, $M_k$, is a conversion from an INLP model~\cite{WieczorekBook}.
The first improvement, $M_{k,all}$, leads to the smallest SAT instances, which are also solved quickly. However, generating this model is too costly. Thus, when problems grow (in terms of $k$, $|S|$, or length of words), $M_{k,all}$ cannot be generated anymore. 
We proposed a set of improvements based on multiset representation of words, $M_{k,mset,l}$. The generated SAT instances are a bit larger with the maximal level than with $M_{k,all}$, but generation is still costly. We thus defined a third improvement based on prefix. On average, the best balance between generation and solving time is obtained with $M_{k,pref}$, $M_{k,mset,1}$, or  $M_{k,mset,3}$: the generation is rather light and the reductions are significant.
The interest of our work is that, to our knowledge, we are the only ones working on CSP model improvements. It is very complicated to compare our results with previous works. Many works on this topics are only formal and experimental results are also difficult to compare.
For examples, the authors of~\cite{jastrzab2016,jastrzab2017} focus on a parallel solver for optimizing $k$. In~\cite{jastrzab2019}, experiments are based on samples issued from the Waltz-DB database~\cite{waltzdb} of amino acid sequences, i.e., all the words are of size 6, and there cannot be any prefix word: in the tests we performed, only anagrams could be used in multisets. Moreover,
%(and thus, there are very few ones, and they make very few reductions of instances). Moreover, 
for all the 50 instances we tried issued from this database, the $M_{k}$ model could be generated and solved in a reasonable time, without need of any model improvement.

In the future, we plan to hybridize $M_{k,mset,l}$ for small values of $l$ with $M_{k,pref}$. 
The second idea is to simplify the work of the SAT solver and of the instance generation with simplified and incomplete training samples. We would then evaluate our SAT models with respect to the accurateness of the generated NFA
%: what is the percentage of positive words (respectively negative words) that are accepted (respectively rejected) from a 
on test set of words.


 \vspace{-3mm}
 \bibliographystyle{splncs04}
 \bibliography{biblio}

\begin{thebibliography}{10}
\providecommand{\url}[1]{\texttt{#1}}
\providecommand{\urlprefix}{URL }
\providecommand{\doi}[1]{https://doi.org/#1}

\bibitem{glucose}
Audemard, G., Simon, L.: Predicting learnt clauses quality in modern {SAT}
  solvers. In: Proc. of {IJCAI} 2009. pp. 399--404 (2009)

\bibitem{waltzdb}
Beerten, J., van Durme, J.J.J., Gallardo, R., Capriotti, E., Serpell, L.C.,
  Rousseau, F., Schymkowitz, J.: {WALTZ-DB:} a benchmark database of
  amyloidogenic hexapeptides. Bioinform.  \textbf{31}(10),  1698--1700 (2015)

\bibitem{delete2}
Denis, F., Lemay, A., Terlutte, A.: Learning regular languages using rfsas.
  Theor. Comput. Sci.  \textbf{313}(2),  267--294 (2004)

\bibitem{DBLP:conf/icgi/Dupont94}
Dupont, P.: Regular grammatical inference from positive and negative samples by
  genetic search: the {GIG} method. In: Proc. of {ICGI} 94. LNCS, vol.~862, pp.
  236--245. Springer (1994)

\bibitem{Garey1979}
Garey, M.R., Johnson, D.S.: Computers and Intractability, {A} Guide to the
  Theory of {NP}-Completeness. W.H. Freeman \& Company, San Francisco (1979)

\bibitem{HeuleMarijn2013Smsu}
Heule, M., Verwer, S.: Software model synthesis using satisfiability solvers.
  Empirical Software Engineering  \textbf{18}(4),  825--856 (2013)

\bibitem{ColinBook}
de~la Higuera, C.: Grammatical Inference: Learning Automata and Grammars.
  Cambridge University Press (2010)

\bibitem{jastrzab2016}
Jastrzab, T.: On parallel induction of nondeterministic finite automata. In:
  Proc. of {ICCS} 2016. Procedia Computer Science, vol.~80, pp. 257--268.
  Elsevier (2016)

\bibitem{jastrzab2017}
Jastrzab, T.: Two parallelization schemes for the induction of nondeterministic
  finite automata on pcs. In: Proc. of {PPAM} 2017. LNCS, vol. 10777, pp.
  279--289. Springer (2017)

\bibitem{jastrzab2019}
Jastrzab, T.: A comparison of selected variable ordering methods for {NFA}
  induction. In: Proc. of {ICCS} 2019. LNCS, vol. 11540, pp. 741--748. Springer
  (2019)

\bibitem{Rossi2006}
Rossi, F., {van Beek}, P., Walsh, T. (eds.): Handbook of Constraint
  Programming. Elsevier Science, 1st edn. (2006)

\bibitem{tomita82}
Tomita, M.: Dynamic construction of finite-state automata from examples using
  hill-climbing. Proc. of the Fourth Annual Conference of the Cognitive Science
  Society pp. 105--108 (1982)

\bibitem{Tseitin1983}
Tseitin, G.S.: On the Complexity of Derivation in Propositional Calculus, pp.
  466--483. Springer Berlin Heidelberg, Berlin, Heidelberg (1983)

\bibitem{DBLP:conf/wia/PargaGR06}
{V{\'a}zquez de Parga}, M., Garc{\'{\i}}a, P., Ruiz, J.: A family of algorithms
  for non deterministic regular languages inference. In: Proc. of {CIAA} 2006.
  LNCS, vol.~4094, pp. 265--274. Springer (2006)

\bibitem{WieczorekBook}
Wieczorek, W.: Grammatical Inference - Algorithms, Routines and Applications,
  Studies in Computational Intelligence, vol.~673. Springer (2017)

\end{thebibliography}

\end{document}